\documentclass[fleqn]{annalen}
\usepackage{graphics}
\begin{document}
\newcommand{\volume}{11}             
\newcommand{\xyear}{2000}            
\newcommand{\issue}{5}               
\newcommand{\recdate}{15 November 1999}  
\newcommand{\revdate}{dd.mm.yyyy}    
\newcommand{\revnum}{0}              
\newcommand{\accdate}{dd.mm.yyyy}    
\newcommand{\coeditor}{ue}           
\newcommand{\firstpage}{507}         
\newcommand{\lastpage}{510}          
\setcounter{page}{\firstpage}        
\newcommand{\keywords}{perfect fluids, general relativity}
\newcommand{\PACS}{04.40.Nr, 04.20.Jb}
\newcommand{\shorttitle}
{Z. Perj\'es, Relativistic perfect fluid models}
\title{Rotating perfect fluid models
in general relativity}
\author{Z.\ Perj\'{e}s}
\newcommand{\address}
{KFKI Research Institute for Nuclear and Particle Physics
\\ Budapest 114, P.O.Box 49, H-1525 Hungary}
\newcommand{\email}{\tt perjes@rmki.kfki.hu  }
\maketitle
 
\begin{abstract}
The various schemes for studying rigidly rotating perfect fluids in general
relativity are reviewed. General conclusions one may draw from
these are: (i) There is a need to restrict the scope of the possible
ans\"atze, and (ii) the angular behaviour is a valuable commodity. This latter
observation follows from a large number of analytic models exhibiting a
NUT-like behaviour. A method of getting around problem (ii) is presented on
a simple example. To alleviate problem (i) for rigidly rotating perfect
fluids, approximation schemes based on a series
expansion in the angular velocity are suggested. A pioneering work, due
to Hartle, explores the global properties of matched space-times to
quadratic order in the angular velocity.
 
As a first example of the applications, it is shown that the
rigidly rotating incompressible fluid cannot be Petrov type D.
\end{abstract}
 
\section{\ Introduction}
 
In the past decades, numerous attempts have been made to construct analytic
models of rotating perfect fluid bodies in general relativity. The static
Schwarzschild metric of an incompressible fluid ball is known since 1916.
The next development, Kerr's rotating solution for the vacuum domain,
took no less than 46 years to happen.
Three decades thereafter, the first, highly idealized model of a
thin rotating disk of dust was found by Neugebauer and
Meinel\cite{Neugebauer and Meinel}. There
are clear indications from this simple limit that the rotating relativistic
star model is a task that is not easy to solve.
 
 A perfect fluid medium in general relativity is characterized by a
stress-energy tensor of the form
\begin{equation}
T_{ik}=\left( \mu + p\right) \mathrm{u}_{i}\mathrm{u}_{k}-pg_{ik}\ ,
\end{equation}
where{\ $p$ is the \textit{pressure},\ $\mu $ the \textit{density }}and {$
\mathrm{u}$}\textrm{\ }{the \textit{four-velocity }}of the fluid{,
normalized by$\mathrm{\ \ u}_{i}\mathrm{u}^{i}=$ 1. }
 
{\ The field equations governing the fluid motion are Einstein's
gravitational equations,
\begin{equation}
R_{ik}-\textstyle\frac{1}{2}g_{ik}R=kT_{ik} \ .
\end{equation}
}
{\ \textit{Barotropic fluids}} additionally obey an{\ equation of state
$p=p\left( \mu \right) . $
The state is called} stationary and axisymmetric when there exist
two commuting\ Killing vectors,
\begin{equation}\textstyle
\xi _{t}=\frac{\partial }{\partial t}\qquad { }\xi _{\varphi }=\frac{
\partial }{\partial \varphi }\ ,
\end{equation}
such that $\xi _{t}$ is time-like and $\xi _{\varphi }$ space-like.
 
The fluid is said to be in \textit{circular rotation} if the
four-velocity $\mathrm{u}$ lies in the plane of the Killing vectors,
\begin{equation}
\mathrm{u}=N\left( \xi _{t}+\Omega \xi _{\varphi }\right) .
\end{equation}
Here \textit{N is a normalization factor} and\ $\Omega $ is \textit{the
angular velocity} of the fluid. \textit{Rigid rotation } is characterized by
the condition that{\textit{\ }$\Omega $ }is a constant, and for differential
rotation: d$\Omega $ $\neq 0$.
 
{\ \textit{Spherical symmetry }}is characterized by a space-time filling
family of{\textit{\ 2-surfaces d}$\vartheta ^{2}+$\textit{\ sin}$
^{2}\vartheta d\varphi ^{2}$ {\ (Delgaty and Lake\cite{Delgaty},
1998).}}
 
{The state is said to possess an \textit{equatorial symmetry }}when it is
invariant under the substitution{\textit{\ }$\vartheta \rightarrow \pi -
 \vartheta .$ }
 
   This paper presents a short review of the subject. The story
begins in Sec. 2, with a Newman-Penrose presentation of the interior
Schwarzschild metric. In Sec. 3, an overview is given of the various
attempts to create more general models which, then, would describe
rotation effects. It will be apparent hence, however, that
success has yet eluded researchers.
The main conclusion to be drawn from this situation is that, in
the interest of better efficiency, there is a need for singling our the
few ans\"atze that correctly encapsulate the physics of the rotating
system. It is suggested here that a very efficient tool for
doing this is given by the series expansion in powers of the angular
velocity
$\Omega$. This scheme has been worked out by Hartle\cite{Hartle} and
used subsequently for a numerical study of stellar
models\cite{Hartle and Thorne, Chandrasekhar and Miller}.
 
\section{The {\textbf{interior Schwarzschild metric }}}
 
The interior Schwarzschild metric, in its pristine form, is written{\ }
\begin{equation}\label{Schw}
ds^{2}=(A-\cos \chi )^{2}dt^{2}-R^{2}[d\chi ^{2}+\sin ^{2}\chi (d\theta
^{2}+\sin ^{2}\theta d\varphi ^{2})] \ ,
\end{equation}
where the radius $R$ of the sphere $S^{3}$ determines the{\ }density,{ {\
$\mu ={3}/{R^{2}}\,$\ }}and the constant { {$ A=3
\sqrt{1-{r_{1}^{2}}/{R^{2}}}$ }}encodes the \textsl{radius }$r_{1}$
\textsl{of the star.}$ $\ The pressure of the fluid is
\begin{equation}
p = \frac{1}{R^{2}}\frac{3\cos \chi -A}{A-\cos \chi }.
\end{equation}
The
familiar radial coordinate $r$ may be reintroduced by writing $\sin \chi =
{r}/{R}.$
 
{\ A Newman-Penrose analysis is conveniently carried out by use of a new,
advanced, time coordinate $u$ such that
\begin{equation}
dt=du+\frac{R}{A-\cos \chi }d\chi .
\end{equation}
}Then the metric takes the form {\
\begin{eqnarray}
ds^{2} &=&(A-\cos \chi )^{2}du^{2}+2R(A-\cos \chi )dud\chi   \label{dsuform}
\\
&&-R^{2} \sin ^{2}\chi (d\theta ^{2}+\sin ^{2}\theta d\varphi ^{2}) \ .
\nonumber
\end{eqnarray}
}
This is in a comoving coordinate system since we have
\begin{equation}
\mathrm{u}=(A-\cos \chi )^{-1}\partial /\partial u \ .
\end{equation}
 
The null tetrad is chosen for the metric (\ref{dsuform}),
\begin{equation}
\ell =\frac{\partial}{\partial \chi} \ \qquad n=\frac{1}{R(A-\cos \chi
)}\frac{\partial}{\partial u}-\frac{1}{2R^{2}}\frac{\partial}{\partial
\chi}
\end{equation}
\[
m=\frac{1}{\sqrt{2}R\sin \chi }\left( \frac{\partial }{\partial \theta }+
\frac{i}{\sin \theta }\frac{\partial }{\partial \varphi }\right)\ .
\]
{\ Thus the four-velocity is given by the linear combination of the tetrad
vectors
\begin{equation}
\textstyle
\mathrm{u}=\frac{1}{2R}\ell + Rn \ .
\end{equation}
}
 
  Inserting in the NP equations, we get the Ricci scalar
\begin{equation}
\Lambda =\frac{1}{4R^{2}}\,\frac{A-2\cos \chi }{A-\cos \chi }
\end{equation}
and{\ the Ricci-tensor components by
\begin{eqnarray}
\Phi _{00} &=&\frac{A}{A-\cos \chi } \qquad
\Phi _{11}  = \frac{A}{4R^{2}(A-\cos \chi )} \qquad
\Phi _{22}  = \frac{1}{4R^{4}}\,\frac{A}{A-\cos \chi } \\
\Phi _{01} &=&\Phi _{02}\ =\ \Phi _{12}\ =\ 0  \ .  \nonumber
\end{eqnarray}
{\ The spin coefficients satisfy the simple relations}
$
\tau =\pi =\overline{\alpha }+\beta =\epsilon -\overline{\epsilon }=0
$.
}
 
The $Weyl$ spinor vanishes,
$\Psi _{K}=0\ $, where $K=0,1,...,4$\ .
The interior Schwarzschild solution is unique among the perfect fluid
metrics in having the property of conformal flatness.
 
{\ {\ $\mathbf{Collinson^{\prime }s\ theorem\ (\cite{Collinson},1976)}:$}}
 
{\ {$\ \mathit{The\ only\ conformally\ flat\ axistationary\
perfect\ fluid\ is\ the\ interior\ }$ \newline}\
{$\mathit{Schwarzschild\ space-time}
\ . $} }
 
\section{{\ \textbf{An (incomplete) compendium of the ans\"atze} }}
 
In this section we review the various schemes deployed for obtaining
stationary and axisymmetric perfect fluid solutions of the gravitational
equations. The entries are arranged in groups characterized by a common
ansatz. For a detailed account of the state of the subject, {\it cf.}
\cite{Chinea}.
 
{\ $\bullet $ \textbf{Petrov type} \newline
\textit{Examples of type D solutions:} \newline
(i) Wahlquist\cite{Wahl}, 1968 solution: This is a rigidly rotating
perfect fluid with the equation of state $\mu + 3p=const.$
\newline
(ii) Senovilla\cite{Senov1}, 1987 solution: The equation of state is
$\mu - p=const $. A later work by the same author, \cite{Senov2} 1992,
presents a wide class of differentially rotating metrics containing a
free function.
\newline \textit{Relation to type N vacuum:} \newline
Hoenselaers \textit{et al.}\cite{Hoenselaers}, 1994 find a map from
the type N vacuum field equations to the equations of rigidly rotating
perfect fluids.
 
{\ $\bullet $ \textbf{Local Rotational Symmetry (LRS)} \newline
Stewart and Ellis\cite{Stewart}, 1968 work out a classification scheme
and find some analytic solutions.
Marklund and Bradley, in a two-part paper \cite{Marklund and Bradley},
1997 and 1999, employ the curvature representation in their study and
obtain examples.}
 
{\ $\bullet $ \textbf{Fluid kinematics} \newline
Chinea and Gonzalez-Romero\cite{CGR}, 1992 develop a
formalism with
differential forms adapted to the kinematics of the fluid}.
\newline
Chinea\cite{Chinea}, 1993 uses this method for getting an analytic
solution with differential rotation, by assuming that the
acceleration of the fluid has a special form. }
 
{\ $\bullet$ \textbf{Killing tensor} \newline
The vacuum Kerr metric possesses a nontrivial Killing tensor. This has
instigated a research of perfect fluid states with a corresponding
property by, among others,
Kramer, Karlovini and Rosquist\cite{KR} and Goliath.
}
 
{\ $\bullet $ \textbf{Vanishing Simon tensor} \newline
\textit{Theorem} (Kramer\cite{Kramer}, 1985): {\ for rigid rotation, the
only possible equation of state is of the form $\mu + 3p=const.$
}\newline Marklund \textit{et al.}\cite{MP}, 1997 generalize the notion
of the Simon tensor for fluids in differential rotation. }
 
{\ $\bullet$ \textbf{Electric-magnetic Weyl}\newline
Rotation is known to give rise to magnetic curvature. Since the interior
Schwarzschild solution has zero conformal curvature, it has been
surmised that the rotating interior Schwarzschild solution will have a
purely magnetic curvature. Fodor \textit{et al.}\cite{FMP},
1999, develop a tetrad formalism in which the electric and
magnetic parts of the curvature explicitly appear. Purely magnetic
curvatures are discussed by Lozanowski and McIntosh \cite{LM}.
 
{\ $\bullet$ \textbf{Lagrangian symmetry} \newline
Stephani\cite{Stephani}, 1988} investigates the invariance
properties of the perfect fluid Lagrangian. He finds that the system
possesses far fewer symmetries than the vacuum Lagrangian.
 
{\ $\bullet$ \textbf{Static$\to$ stationary symmetry} \newline
Herlt\cite{Herlt}, 1988 elaborates on methods by which to a given
static state a rotating counterpart can be obtained.}
 
{\ $\bullet $ \textbf{Geodesic eigenrays} \newline
Lukacs et al\cite{Lukacs}, 1983, obtain a fluid
solution with geodesic eigenrays, exhibiting NUT-like behaviour.}
 
   From this impressive array of theoretical efforts, and from
the fact that the only global solution properly matched to the ambient
vacuum region is the infinitesimally thin,  flat disk of dust of
Neugebauer and Meinel\cite{Neugebauer and Meinel},
it transpires that {\em the right ansatz has
not yet been found.}
 
Hence we may draw two important conclusions for this research:
 
{\ \textbf{CONCLUSION 1}: There is a clear need to somehow restrict the
scope of possible ans\"atze.} The suggestion here is to resort to
perturbation techniques. In the next section it will be shown that the
approximation scheme developed originally by Hartle \cite{Hartle} is an
efficient tool for selecting the physically acceptable schemes.
 
From the relatively large number of rotating fields with unacceptable
causal behaviour, we draw
 
{\ \textbf{CONCLUSION 2}: The angular behaviour
is a valuable commodity.}

\section{Linear perturbations}
In this section we demonstrate the advantages of
a perturbative treatment in the slow-rotation
limit. Although previous works focussed mainly on numerical
applications, we want to point out that perturbative methods can be
used here like a torchlight when trying to find
the causes of the difficulties met with analytic models.
 
\subsection*{(i) The lesson from the vacuum}
 
Rotating vacuum fields are governed by the Ernst equation
\cite{Ernst}, (1968):
 
\begin{equation}
\left( \xi \overline{\xi }-1\right) \Delta \xi =2\overline{\xi }\nabla \xi
\cdot \nabla \xi\ .
\end{equation}
In spheroidal coordinates, this is written
 
\begin{eqnarray}
\left( \xi \overline{\xi } - 1 \right)&&\!\!\!\!\!\!\!\!\!
\left[ \frac \partial {\partial x}\left( x^2-1\right) \frac \partial
{\partial x}-\frac \partial {\partial y}\left( 1-y^2\right) \frac \partial
{\partial y}\right] \xi \\
&&= 2\overline{\xi }\left[ \left( x^2-1\right) \left( \frac{\partial \xi }{
\partial x}\right) ^2-\left( 1-y^2\right) \left( \frac{\partial \xi }{
\partial y}\right) ^2\right] \ .\nonumber
\end{eqnarray}
 
Linearizing around the Schwarzschild solution $\xi =x$,
\begin{equation}
\xi =x+i\xi _1\qquad \left( \xi _1\mathrm{\ real}\right)
\end{equation}
the derivative of the first-order function $\xi_1$ satisfies the
Laplace equation:
\begin{equation}
\Delta \frac{\partial ^2\xi _1}{\partial x^2}=0\ .
\end{equation}
 
{\  {Hence
\begin{equation}
\frac{\partial ^2\xi _1}{\partial x^2}=\sum_{l=2}^\infty a_lQ_l\left(
x\right) P_l\left( y\right)
\end{equation}
} }
where $P_l\left( y\right)$ is the Legendre polynomial of order $l$
and
\begin{equation} \label{xi1}
\xi _1=\sum_{l=2}^\infty a_l\int \left( \int Q_l\left( x\right) dx\right)
dxP_l\left( y\right) +ay \ .
\end{equation}
 
 The analytic solutions of the Ernst equation found by Tomimatsu and
Sato (\cite{TS}, 1972) have the rational form
 
\begin{equation}
\xi =\alpha /\beta  \ .
\end{equation}
 
Fortuitously, the integral (\ref{xi1}) is of polynomial character.
The polynomials in $\xi _1$ have been used for the leading terms in
$\alpha
.$
 
\subsection*{(ii) Perfect fluids}
 
\textit{Hartle } \cite{Hartle} (1967) casts the
metric of a slowly and rigidly rotating perfect fluid in the following
form:
\begin{eqnarray}
ds^2 &=&e^\nu dt^2-e^\lambda dr^2-r^2\left[ d\vartheta ^2+\sin ^2\vartheta
\left( d\varphi -\omega dt\right) ^2\right]
+\mathcal{O}\left( \Omega ^2\right) \ .
\end{eqnarray}
From the large-$r$ behaviour it follows that the rotational potential
$\omega$ is a function of the radius alone, $\omega =\omega \left(
r\right) $.
 
From the Schwarzschild solution (\ref{Schw}) for an
$\textit{incompressible fluid}$ we get:
 
\begin{equation}
e^\nu =(A-\cos \chi )^2\ ,\qquad e^\lambda =\frac{R^2}{\sin ^2\chi }\ .
\end{equation}
 
{\ The potential {$\omega $ \textit{satisfies Heun's equation}
\begin{equation}\label{Heun}
(A-z)(z^2-1)\frac{d^2\omega}{dz^2} -(3z^2-5Az+2)\frac{d\omega}{dz}
+4A\omega=0
\end{equation}
with $z=\cos\chi$ \ .} }
 
This equation has proved to be a stumbling block in the way of progress
for three
decades. Na\"\i ve attempts at removal do fail. For example, one may ask
for the effect of coordinate change. Since the $\omega$ equation is a
special case of the six-parameter
Heun Eq., it would seem conceivable that in a better coordinate system,
solution is more amenable. Such a new coordinate system is provided, for
example, by the property of the
interior Schwarzschild metric that it is conformally flat. As a
consequence, the metric can be conformally mapped to the Einstein
Universe (as the case is with the Minkowski spacetime):
 
\begin{equation}\label{confsch}
ds^2=\Omega^2\left[dt^2-dx^2-\sin^2x(d\vartheta^2+\sin^2\vartheta
d\varphi^2)\right]\ .
\end{equation}
The conformal factor is $ \Omega^2=R^2(A^2-1)/(A-\cos x)^2$.
The transformation to the conformal form (\ref{confsch}) is as follows,
$\ \sin x= (A^2-1)^{1/2}(A-\cos\chi)^{-1} \sin\chi $\ .
We find, however, that the perturbation equation is of the Heun type
again.
 
 In a like manner, attempts to change the dependent variable have failed
to turn up a solvable equation. A straightforward thought here is to try
and use the Ernst potential in place of the rotation potential $\omega$
in the metric
\begin{eqnarray}
ds^2 &=&f\left( dt+\omega d\varphi \right) ^2 \\
&&-f^{-1}\left[ e^{2\gamma }dr^2 +\left( 1-r^2\right) \left(
(1-y^2)^{-1}dy^2
+\left( 1-y^2\right) d\varphi ^2\right) \right]\ ,\nonumber
\end{eqnarray}
where
$
y=\cos \vartheta
$.
The field equation for the Ernst potential $\xi$ is
\begin{equation}
\textstyle
\left( \xi \overline{\xi }-1\right) \Delta \xi -2\overline{\xi }\nabla \xi
\cdot \nabla \xi =\frac 12k\left( \xi \overline{\xi }-1\right) \left( \xi
+1\right) ^2\left( \mu +3p\right)
\end{equation}
with $p$ the pressure and $\mu$ the density of the fluid.
 
Integration of the Bianchi identity yields
\begin{equation}
\left( \mu +p\right) f^{1/2}=const.
\end{equation}
For the interior Schwarzschild metric, $\gamma =0,$ and
 
\begin{equation}
f=\left( A-r\right) ^2,\qquad k\mu =3,\qquad kp=\frac{3r-A}{A-r}\ .
\end{equation}
 
Let the rotational perturbation $\xi_1$ be introduced,
\begin{equation}
\xi =\frac{1+f}{1-f}+i\xi _1 \ .
\end{equation}
{\  {The linearized field equation for $\xi _1$ is
separable; the $y$ dependence satisfies the Legendre equation of index
$l$. } }
 
{\  {It is possible to integrate the radial equation for $l=0$:
\[ \!\!\!\!
\xi _1=\frac 1{\left[ (r-A)^2-1\right] ^2}\!\left\{\! C_1
  - C_2\!\left[ \frac{r^2\!\!+\!3Ar\!+\!A^3r\!-\!2\!-\!3A^2}{\sqrt{r^2-1}}
+\!3A\ln \left( r+\!\sqrt{r^2-1}\right) \right] \!\right\}.
\]
} }
\noindent This would appear quite encouraging.
However, from asymptotic matching, one finds that $\xi _1$ is
proportional to $y$, and hence we need the integral for $ l=1$,
which is not known in a closed form.
 
\section{Two theorems}
 
In this section we describe how to restrict the scope of choice in
order to meet
 Conclusion 1, by applying the perturbative treatment to various
systems.
 
\subsection{Theorem on incompressible fluids}
 {\ To second order in $\Omega $, the metric of the rigidly rotationg
incompressible fluid is\cite{Hartle} \begin{eqnarray}
ds^2 &=&(1+2h)(A-\cos \chi )^2dt^2
-R^2(1+2k)\sin ^2\chi \sin ^2\vartheta (d\varphi -\omega dt)^2 \nonumber\\
&&-R^2(1+2m)d\chi ^2-R^2(1+2n)\sin ^2\chi d\vartheta ^2\ .
\end{eqnarray}
Here $\omega $ is of first order, and\ $h,$ $k,$ $m,$ $n$ of second
order in the angular velocity $\Omega$.}
 
{\ A comoving tetrad may be chosen } {\
\begin{eqnarray}
&&e_0^t=\frac{1-h}{A-\cos \chi }+(\omega -\Omega )^2\frac{R^2\sin
^2\vartheta \sin ^2\chi }{2\left( A-\cos \chi \right) ^3}\nonumber  \\
&&e_0^\varphi =\frac \Omega {A-\cos \chi }\ ,\quad\qquad
e_1^\chi =\frac{1-m}R\  \\
&&e_2^\vartheta =\frac{1-n}{R\sin \chi }\ ,\qquad\qquad
\ e_3^t=\frac{R\sin \vartheta
\sin \chi (\omega -\Omega )}{\left( A-\cos \chi \right) ^2}\nonumber  \\
&&e_3^\varphi =\frac{k-1}{R\sin \vartheta \sin \chi }+\left( \omega
^2-\Omega ^2\right) \frac{R\sin \vartheta \sin \chi }{2\left( A-\cos \chi
\right) ^2}\ .\nonumber
\end{eqnarray}
The vorticity is given by the first-order rotation coefficients
\begin{eqnarray}
\omega _1 &=&\frac 1{2\left( A-\cos \chi \right) }\left[ 2\cos \vartheta
(\Omega -\omega )-\sin \vartheta \frac{\partial \omega }{\partial \vartheta }
\right] \\
\omega _2 &=&\frac{\sin \vartheta }{2\left( A-\cos \chi \right) ^2}\left[
2(\omega -\Omega )(A\cos \chi -1)
 +\sin \chi \left( A-\cos \chi \right)
\frac{\partial \omega }{\partial \chi }\right]\ .
\end{eqnarray}
}
The changes in the other rotation coefficients are of
the second order.
 
The magnetic part H$_i$ of the Weyl tensor is of order $\Omega $,
{\
\begin{eqnarray}
H_1 &=&\frac{\cos \vartheta }{R\left( \cos \chi -A\right) }\frac{d\omega }{
d\chi }\ ,\qquad \ H_2=-\frac 12H_1 \\
H_3 &=&-\frac{\omega _2\cos \chi }{R\sin \chi }+\frac{\sin \vartheta }{R\sin
\chi }\frac{\omega -\Omega }{A-\cos \chi }\nonumber\ .
\end{eqnarray}
The electric part $E_i$ is of order $\Omega ^2$.
The condition for the space-time to be Petrov type II, to quadratic
order, is $H_3=0$ and $H_1=H_2.$ Hence the Petrov type of this
incompressible fluid cannot be II.
 
The Petrov type is D if $H_3^2-2H_1^2-2H_2^2-5H_1H_2=0.$
The general solution of this condition has the form
\begin{equation}
\omega =C\left(  A/{\cos \chi }-1\right) ^2\ .
\end{equation}
This is not a solution of the Heun equation (\ref{Heun}) which proves
the following
 
{\ \textbf{THEOREM\cite{FP}:} \ A rigidly rotating incompressible fluid
ball in an asymptotically flat vacuum exterior cannot be Petrov type D. }
 
\subsection{Theorem on the Wahlquist solution}
The Wahlquist metric is an example of type D solutions. It is given
by\cite{Wahl} {\
\begin{eqnarray}
ds^2 &=&f\left( dt-Ad\varphi \right) ^2-r_0^{\ 2}\left( \zeta ^2+\xi
^2\right) \left[ \frac{d\zeta ^2}{\left( 1-\mathrm{k}^2\zeta ^2\right) h_1}
\right. \\
&&\left. \qquad \qquad \quad \quad +\frac{d\xi ^2}{\left( 1+\mathrm{k}^2\xi
^2\right) h_2}+\frac{h_1h_2}{h_1-h_2}d\varphi ^2\right] \ ,\nonumber
\end{eqnarray}
where
\begin{eqnarray}
f &=&\frac{h_1-h_2}{\zeta ^2+\xi ^2}\quad \qquad A=r_0\left( \frac{\xi
^2h_1+\zeta ^2h_2}{h_1-h_2}-\xi _A^{\ 2}\right) \\
h_1\left( \zeta \right) &=&1+\zeta ^2+\frac \zeta {\kappa ^2}\left[ \zeta
-\frac 1{\mathrm{k}}\sqrt{1-\mathrm{k}^2\zeta ^2}\arcsin \left( \mathrm{k}
\zeta \right) \right] \\
h_2\left( \xi \right) &=&1-\xi ^2-\frac \xi {\kappa ^2}\left[ \xi -\frac 1{
\mathrm{k}}\sqrt{1+\mathrm{k}^2\xi ^2}\textrm{arcsinh}\left(
\mathrm{k}\xi \right) \right] \ .
\end{eqnarray}
The constant $\xi _A$ is defined by $h_2\left( \xi _A\right) =0$ . }
 
\ The pressure and density are
\begin{equation}
p=\textstyle\frac 12\mu _0\left( 1-\kappa ^2f\right) \quad \qquad \mu =\textstyle\frac 12\mu
_0\left( 3\kappa ^2f-1\right)  \ .
\end{equation}
The equation of state is
\begin{equation}
\mu _0=\mu +3p=\frac{2\mathrm{k}^2}{\kappa ^2r_0^{\ 2}} \ .
\end{equation}
 
{\textbf{THEOREM\cite{BFMP}:} By quadratic approximation in $\Omega $}
{\ , }
 
{\  \emph{The Wahlquist metric cannot be matched to an asymptotically
flat vacuum exterior}}.
 
\section{\ How to secure the angular behaviour?}
 I do not really have a satisfactory answer to
{\it  Conclusion 2} concerning the angular behaviour. All I can offer
here is a particular
{{recipe} by which one creates a relativistically rotating
version of a certain static system. There is no
guarantee, however, that the space-time so obtained will have the
desired type of matter source. Succinctly stated, the idea
is as follows. The constant-time
sections of the Schwarzschild fluid ball (\ref{Schw}) are $S^3$
spheres. Now an $S^3$ may be squeezed
and simultaneously set in rotation about the $z$ axis following these
steps:
 \newline
\noindent
\textbf{--} Embed {$\mathcal{S}^3$} in $E^4$ with the metric
\begin{equation}
ds^2=(dx^1)^2+(dx^2)^2+(dx^3)^2+(dx^4)^2\ .
\end{equation}
Introduce polar coordinates $(R,\chi,\vartheta,\varphi)$,
\begin{eqnarray}
x^1+ix^2=R\sin \chi \sin \vartheta e^{i\varphi }  \nonumber \\
x^3=R\sin \chi \cos \vartheta   \\
x^4=R\cos \chi  \ .\nonumber
\end{eqnarray}
The surface {$\mathcal{S}^3$} is given by $R=const.$ } }
 
\noindent
{\  {\textbf{--} Squeeze {$\mathcal{S}^3$} and set simultaneously in
rotation by putting
\begin{equation}
\chi\rightarrow\chi+ia
\end{equation}
in $x^1+ix^2$. } }
{\  {Thus the $R=const.$ surface is flattened, and has the metric } }
\begin{eqnarray}
ds^2&=& (\sin^2\chi+\sinh^2 a\cos^2\vartheta)(d\vartheta^2+\sin^2\vartheta
d\varphi^2)\   \nonumber \\
&+&\sinh a \cosh a\sin^2 d\chi d\varphi - \sinh^2 a \sin^4 \vartheta
d\varphi^2 \ . \nonumber
\end{eqnarray}
 
\noindent
{\  {\textbf{--} Embed $\mathcal{S}^3$ in the Einstein Universe.} }
For $R=1$, the metric is of the form
\begin{equation}
ds^2=dt^2-d\chi^2-\sin^2\chi(d\vartheta^2+\sin^2\vartheta d\varphi^2)\ .
\end{equation}
 
\noindent
{\  {\textbf{--} Soak up the {\ slack $d\chi \,d\varphi $ term by
introducing the advanced time coordinate}
\begin{equation}
u=t+\chi \ .
\end{equation}
} }
 
{\  {The space-time metric then takes the form:
\begin{eqnarray}
ds^2 &=&\!\!\left( du-\sinh a\sin ^2\vartheta d\varphi \right) ^2
\nonumber \\
-\!\! &2&\!\!\left( du-\sinh a\sin ^2\vartheta d\varphi \right) \left( \cosh ad\chi
-\sinh a\sin ^2\vartheta d\varphi \right)  \nonumber \\
&-&\!\!(\sin ^2\chi +\sinh ^2a\cos ^2\vartheta )(d\vartheta ^2+\sin ^2\vartheta
d\varphi ^2)  \ .\nonumber
\end{eqnarray}
} }
{\  {This metric represents a squeezed and rotated Einstein universe.
In the limit as $R\to\infty$, we recover Minkowski space-time. } }
 
\textbf{CONJECTURE }: By continuity in the rotation
parameter $a$,      \newline
{\  {\emph{There is an open neighbourhood of Minkowski space-time in
the minisuperspace of squeezed Einstein universes that is causally
well-behaved}. } }
 
 Application of this procedure to the interior Schwarzschild metric
yields rotating models with various stress-energy tensors, depending on
details of the embedding.
 
\vspace*{0.25cm} \baselineskip=10pt{\small \noindent
It is a pleasure to thank the Organizers of the
\textit{Journees Relativistes} Conference for providing a seamless
platform for scientific activity and for hospitality. Part of this work
has been done under the contract OTKA T022533.}

\end{document}